\documentstyle[12pt,epsf]{article}
\newcommand{\be}{\begin{equation}}
\newcommand{\ee}{\end{equation}}
\newcommand{\bea}{\begin{eqnarray}}
\newcommand{\eea}{\end{eqnarray}}
\newcommand{\bdm}{\begin{displaymath}}
\newcommand{\edm}{\end{displaymath}}
\begin{document}
\large{

\title{ Hysteresis in One-Dimensional Anti-Ferromagnetic Random-Field 
Ising Model at Zero-Temperature }

\author{  Prabodh Shukla, Ratnadeep Roy, and Emilia Ray
\\Physics Department \\ North Eastern Hill University
\\ Shillong-793 022, INDIA }
\maketitle
\begin{abstract}

We analyse hysteresis in a one-dimensional anti-ferromagnetic random field
Ising model at zero-temperature. The random field is taken to have a
rectangular distribution of width $2 \Delta$ centered about the origin. A
uniform applied field is varied slowly from $-\infty$ to $+\infty$ and
back. Analytic expression for the hysteresis loop is obtained in the case
$\Delta \le |J|$, where $|J|$ is the strength of the nearest neighbor
interaction.

\end{abstract}

\vspace{3cm}
email: shukla@dte.vsnl.net.in
\clearpage
\section{Introduction}

In a recent paper~\cite{shukla1} , we presented an analytic solution of
the zero-temperature dynamics of the anti-ferromagnetic random field Ising
model in a slowly varying uniform applied field over a truncated range of
the field. The purpose of the present paper is to extend the earlier
solution to cover the entire range of the applied field, and hence to
obtain an analytic expression for the hysteresis loop. We refer the reader
to reference [1] for a detailed discussion of the model, and the method of
its solution. Here, we recapitulate the salient points of reference [1]
for maintaining continuity with it, and to set up the notation.

The Hamiltonian of the system is given by
\be
H=-J \sum_{i}{s_{i}s_{i+1}}-\sum_{i}h_{i}s_{i}-h_{a}\sum s_{i},
\ee

where $ s_{i}=\pm{1}$ are Ising spins at sites $i=1,2,3 \ldots$ of a
one-dimensional lattice, $h_{a}$ is a uniform applied field, and $h_{i}$
is a quenched random field drawn independently from a continuous bounded
distribution,

\begin{eqnarray}
p(h_{i}) & = \frac{1}{2\Delta} &\qquad\mbox{if}\qquad
-\Delta \le h_{i} \le \Delta
\nonumber \\ & = 0, & \qquad\mbox{    otherwise.}
\end{eqnarray}

The external field $h_{a}$ is slowly increased from $-\infty$ to
$+\infty$, and the system is allowed to relax by the single-spin-flip
Glauber dynamics at zero temperature, i.e. spins are allowed to flip one
at a time if it lowers their energy. This dynamics takes the system to a
stable state where each spin has the same sign as the net local field
$l_{i}$ at its site.

\be s_{i}=\mbox{ sign }{l_{i}} =\mbox{ sign }
[J(s_{i-1}+s_{i+1})+h_{i}+h_{a}] \ee

The magnetization $m(h_{a})$ per spin at the applied field $h_{a}$ is
given by

\be
m(h_{a})=\frac{1}{N}\sum_{i}s_{i}
\ee

Our object is to calculate $m(h_{a})$ for all values of $h_{a}$ in the
range $[-\infty$ to $+\infty]$. In reference [1], we obtained $m(h_{a})$
for $[-\infty \le h_{a} \le 2|J|-\Delta]$ in the case $[\Delta \le |J|]$.
The result is expressed conveniently in terms of variables h and p defined
as follows:

\bdm
h=-h_{a} \mbox{   mod } |2J| ; \mbox{    }
p(h)=\int_{h}^{\Delta}p(h_{i})dh_{i} = \frac{\Delta-h}{2\Delta}  
\edm

The field h serves as a reduced applied field whose importance is
explained below. If $\Delta \le |J|$, the spins turn up in three
categories. The first to turn up are the spins whose nearest neighbors are
both down (ramp-I), then the spins with one neighbor up and one down
(ramp-II), and finally the spins with both neighbors up (ramp-III).  On
each ramp, a spin with quenched field $h_{i}$ turns up when the reduced
field h crosses the threshold $h+h_{i}=0$. Thus the reduced field acts as
an effective applied field on each ramp. The reduced field covers the same
range as the quenched field, i.e. $[ -\Delta \le (h,h_{i}) \le \Delta ]$.
The parameter p is the fraction of sites in the system with $h_{i} \ge h$.
This parameter also serves as a convenient measure of the reduced field on
each ramp. A key quantity on each ramp is the fraction of spins whose
quenched field is larger than the reduced field h. This quantity
determines the shape of the ramp. The difficulty in calculating this
quantity comes from the fact that the distribution of quenched fields on
each ramp is modified a posteriori by the dynamics of the system. In
reference [1], we obtained the following results:\\

\noindent Ramp-I: \hspace{.25cm}$ [ -2|J| -\Delta \le h_{a} \le
-2|J|+\Delta)] $ \\ \bdm m(h)  = -e^{-2p(h)} \edm \\ Plateau-I:
\hspace{.25cm}$ [ -2|J|+\Delta \le h_{a} \le -\Delta ]$ \bdm m(h) = e^{-2}
\edm \\ Ramp-II: \hspace{.25cm} $ [ -\Delta \le h_{a} \le \Delta ] $ \bea
m(h) = - \left( 3 + 8 e^{-1} +e^{-2} \right)  + 8 \left( 1 + e^{-1}
\right) p - 4 p^{2} \nonumber \\ -\frac{2}{3} {\left( 1 + e^{-1}
\right)}^{2} p^{3} + \frac{5}{6} \left( 1 + e^{-1} \right) p^{4} -
\frac{4}{15} p^{5} \nonumber \\ + \left\{ 8 \left( 1 + e^{-1} \right) - 8
p -2 \left( 1 + e^{-1} \right) p^{2} + \frac{4}{3} p^{3} \right\} e^{-p}
\nonumber \\ - \left(5 + 2 p\right) e^{-2p} \nonumber \eea \\ Plateau-II:
\hspace{.25cm} $ [ \Delta \le h_{a} \le 2|J|-\Delta ] $ \bdm m(h) = \left[
\frac{27}{30} -\frac{7}{6} e^{-1} -\frac{8}{3} e^{-2} \right] \\ \edm

Reference [1] did not contain results for ramp-III, i.e. $m(h_{a})$ in the
range $ [ 2|J|-\Delta \le h_{a} \le 2|J|+\Delta ] $. This calculation is
taken up in the following.

\section{The Ground State}

Plateau-II forms the ground state for the evolution of ramp-III in an
increasing applied field. The ground state has solitary down spins
dispersed in a sea of up spins. Following the nomenclature of reference
[1], we call the down spins singlets. The net field on a singlet is equal
to $-2|J| + h_{i} + h_{a}$. This lies in the range $[-2|J| - \Delta +
h_{a}]$ to $[-2|J| + \Delta + h_{a}]$. Therefore, the singlets begin to
turn up at $h_{a}=2|J|-\Delta$, and are all up at $h_{a}=2|J|+\Delta$.
These limits mark the beginning and the end of ramp-III. The shape of
ramp-III is determined by the fraction of up spins at applied fields
between these limits. We take an arbitrary value of the applied field
$h_{a}=-2|J| -h$ on ramp-III, and calculate the fraction of singlets which
are up at this field. In other words, we calculate the fraction of
singlets with $h_{i} \ge h$ or equivalently $p_{i} \le p$, where

\bdm
p_{i}=\frac{\Delta-h_{i}}{2\Delta}.
\edm 

As discussed in reference [1], the singlets on plateau-I fall into three
categories. In category (1) are the singlets which were created on ramp-I.
The fraction of singlets which belong to this category, and whose quenched
field is greater than h is given by,

\be
P^{II}_{1}(p)=p-\frac{1}{2}\left[1-e^{-2p}\right]
-\frac{2}{e}\left[e^{-p}-(1-p)\right]
\ee

In category (2) are the singlets which were created on ramp-II by a
vanishing doublet. The fraction of singlets in this category whose
quenched field is greater than h is equal to,

\be
P^{II}_{2}(p)=\left[e^{-p}-(1-p)\right]^2
\ee

Finally, category (3) singlets are those that were created on ramp-II by
the central spin of an unstable up triplet flipping down. The fraction of
singlets in this category whose quenched field is larger than h is given
by,

\bea
P^{II}_{3}(p)= -\left( 1 + 2 e^{-1} \right)
+ \left[ \frac{1}{4} + 2 e^{-1} + 4 e^{-2} \right] p & & \nonumber \\
- \frac{1}{2} \left[ 1 + 5 e^{-1} + 4 e^{-2} \right] p^2 
+ \frac{1}{3} \left[ \frac{3}{2} 
+ 4 e^{-1} + e^{-2} \right] p^3 & & \nonumber \\
-\frac{1}{4} \left[ 1+ e^{-1} \right] p^4 + \frac{1}{20} p^5  + \left[1+ 2
e^{-1}\right]e^{-p}  & & \nonumber \\ 
-2 p e^{-(1+p)} +p^2 e^{-p}
+\frac{1}{2} \left[ 1-e^{-2p} \right] & & 
\eea

The total fraction of singlets present on plateau-II whose quenched fields
is larger than h is equal to,

\be 
P^{II}(p)= P^{II}_{1}(p) + P^{II}_{2}(p) + P^{II}_{3}(p)
\ee

The above equation gives the fraction of singlets which will turn up in an
applied field $h_{a}=2|J|-h$ from among the singlets initially present on
plateau-II. However, when these singlets turn up, some of their nearest
neighbors turn down. This process creates new singlets~\cite{ns}. The
newly created singlets turn up later at a higher applied field on
ramp-III. We need to calculate the fraction of newly created singlets, and
their restoration to the original state as function of the applied field
before we can determine ramp-III.


\section{Creation of New Singlets}

In this section, we consider the circumstances in which the destruction of
a singlet on ramp-III accompanies the creation of a new singlet. Consider
a singlet in the ground state, say the spin at site 3 in Figure 1. Its
nearest neighbors at sites 2 and 4 are up. A next nearest neighbor at site
1 is down, and the other next nearest neighbour at site 5 is up. Suppose
the singlet just flips up on ramp-III at an applied field $h_{a}$, i.e.
$h_{3} - 2 |J| + h_{a} = \epsilon$, where $\epsilon \ge 0$. The applied
field at this point is $h_{a} =2 |J| - h_{3} + \epsilon$. We ask the
question, could a nearest neighbor of the singlet flip down when the
singlet flips up? 

First, consider the nearest neighbor at site 2 in Figure 1. After site 3
has flipped up, site 2 has one nearest neighbor up and one down. Thus site
2 may flip down if $h_{2} + h_{a} \le 0$, or $h_{2} \le h_{3} - 2|J|
-\epsilon$. However, this is not possible because $h_{2}$ and $h_{3}$ lie
in the range [$-\Delta$ to +$\Delta$], and we are considering the case
$\Delta \le |J|$. Thus a nearest neighbor of a singlet which has both its
nearest neighbors down will stay up when the singlet turns up as long as
$\Delta \le |J|$.

Next, consider the spin at site 4. After the singlet has turned up, site 4
has both its nearest neighbors up. It will turn down if $-2|J| + h_{4} +
h_{a} \le 0$, or $h_{4} \le h_{3} - \epsilon$. If $h_{4} \le h_{3}$, site
4 will turn down when site 3 turns up. After site 4 has turned down, site
3 has one neighbor up and one down. The net field at site 3 is then $h_{3}
+ h_{a} = 2|J| +\epsilon$, which is positive. Thus site 3 will stay up
after site 4 has turned down.  We conclude that, when a singlet turns up,
its nearest neighbor may turn down if that nearest neighbor has less
quenched field than the singlet, and also if it had one nearest neighbor
already up before the singlet turned up. Note that when site 4 turns down,
it increases the upward field at site 5. There is no possibility of site 5
turning down as a result of site 4 turning down. Consequently, a spin
turning up on ramp-III does not cause any change in the state of spins
beyond the nearest neighbors (absence of avalanches).

Before proceeding further, we rewrite the two rules which will guide us in
the following analysis. \begin{enumerate} \item[Rule-1:] When a singlet
turns up on ramp-III, its nearest neighbor stays up if the adjacent next
nearest neighbor is down, and $\Delta \le |J|$. \item[Rule-2:] When a
singlet turns up on ramp-III, its nearest neighbor turns down if both of
the following conditions are satisfied:\begin{enumerate}\item[(a.)] the
adjacent next nearest neighbor is up, and \item[(b.)] the nearest neighbor
has less quenched field than the singlet.\end{enumerate}\end{enumerate}

The next question is, in what circumstances, the quenched field on a
nearest neighbor of a singlet on plateau-II can be smaller than the
quenched field on the singlet. We examine various possible cases.  
Suppose the singlet in question was created on ramp-II by the destruction
of a doublet. For example, look at Figure 1 again and suppose sites 2 and
5 were up, and 3 and 4 were down on plateau-I. This requires $h_{3} \le
h_{2}$ and $h_{4} \le h_{5}$. Let $h_{3} \le h_{4}$ without any loss of
generality. In this case, site 4 will turn up on ramp-II. When site 3
turns up on ramp-III, no new singlet can be created because both nearest
neighbors of site 3 have a larger quenched field than site 3 (Rule-2).

Next, let us assume that the singlet in question was created on ramp-II by
the unstable central spin of an up triplet flipping down. The central spin
of an up triplet flips up for the the first time on ramp-I. It separates
two adjacent doublets on plateau-I. Therefore, the quenched field on the
central spin is larger than the quenched field on each of its nearest
neighbors. The nearest neighbors of the central spin flip up on ramp-II
(the central spin flips down at this event). Therefore each nearest
neighbor of the central spin has a larger quenched field than the next
nearest neighbor of the central spin which is adjacent to it. Consequently
the central spin has a larger quenched field than each of its next nearest
neighbors. When the central spin flips up for the second time on ramp-III,
its next nearest neighbors are still down. Therefore there is no chance
(Rule-2) for the nearest neighbors of the central spin to flip down when
the central spin flips up on ramp-III.


Having ruled that the destruction of a singlet created on ramp-II does not
give rise to a new singlet on ramp-III, we are left with singlets created
on ramp-I and present on plateau-I. When one of these singlets disappears
on ramp-III, it may create a new singlet if one of its nearest neighbors
has a smaller quenched field than the singlet, and if the next nearest
neighbor is up. We note that only one of the two nearest neighbors of a
singlet on plateau-I can have a quenched field which is smaller than the
singlet. A site which has a larger quenched field than both of its nearest
neighbors must be up on plateau-I. Now consider a singlet on plateau-I as
shown at site 2 in Figure 2. Let its nearest neighbor on the right (at
site 3) have a smaller quenched field ($h_{3} \le h_{2}$). Site 4 must be
down because there are no strings of up spins of length greater than unity
on plateau-I. Site 5 may be up or down. The two possibilities for site 5
are depicted in Figures 2 and 3 respectively. In Figure 2, a singlet is
followed by a singlet on plateau-I. In Figure 3, a singlet is followed by a
doublet.

Consider Figure 2 first. What is the probability per site that such an
object occurs on plateau-I? A singlet on plateau-I must be followed by a
singlet or a doublet. It was shown in reference [1], that the probability
per site of the occurrence of a doublet is equal to $e^{-2}$, and the
probability that a doublet is followed by a doublet is equal to
$\frac{1}{3} e^{-2}$. Therefore, the probability that a doublet is
followed by a singlet (or vice-versa) is equal to $\frac{2}{3} e^{-2}$. It
was also shown that the probability of the occurrence of a singlet is
equal to $\frac{1}{2} \left[ 1 - 3e^{-2} \right]$. Therefore, the
probability per site that a singlet is followed by a singlet on plateau-I
is equal to $\frac{1}{2} \left[ 1 - \frac{13}{3}e^{-2} \right]$.

We have obtained above the probability of the occurrence of objects shown
in Figure 2. Our immediate interest lies in a subset of these objects with
$h_{3} \le \mbox{ min } (h_{2},h_{4}) $. This subset determines the
creation of new singlets. Suppose $h_{2} \le h_{4}$. Then on ramp-III,
site 4 will flip up before site 2. When site 2 flips up, site 3 will flip
down because the conditions for the creation of a new singlet are
fulfilled (Rule-2). In order to calculate the fraction of newly created
singlets, we need to know the distribution of fields $h_{2}$, $h_{3}$, and
$h_{4}$. These may be obtained on the lines of reference [1] if one notes
that sites 1 and 5 must have flipped up on ramp-I before site 3. Site 1
must have flipped up before site 3 because $h_{1} \ge h_{2} \ge h_{3}$.
Similarly, site 5 must have flipped up before site 3 because $h_{5} \ge
h_{4}$, $h_{4} \ge h_{2}$, $h_{2} \ge h_{3}$, and therefore $h_{5} \ge
h_{3}$. This proves that just before site 3 flipped up on ramp-I, sites 2,
3, and 4 formed a string of three down spins bordered by up spins at sites
1 and 5. The screening property discussed in reference [1] can be applied
here to conclude that the distributions of $h_{2}$, and $h_{4}$ are
independent of each other, and each is given by

\bdm
Prob[p \le p_{2} \le p+dp] =[1-e^{-p}] dp  
\edm
\bdm
Prob[p \le p_{4} \le p+dp] =[1-e^{-p}] dp
\edm
where
\bdm
p_{i}=\frac{\Delta-h_{i}}{2\Delta}
\edm

Given that $h_{2}$ lies in the range [h to h+dh], $h_{3}$ must be
uniformly distributed in the range [$-\Delta$ to h]. Thus,

\bdm
Prob[p \le p_{3} \le p+dp] = dp \mbox{    } (\mbox{if   }p_{3} \ge
p_{2})
\edm

The contribution of Figure 2 to the fraction of newly created singlets
when all sites with $h_{i} \ge h$ have been relaxed on ramp-III is given
by,

\bea
P^{III}_{4}(p) = 2 \int_{0}^{p} \left[ 1-e^{-p_{2}} \right] dp_{2}
\int_{p_{2}}^{1}
dp_{3} \int_{0}^{p} \left[ 1 -e^{-p_{4}} \right] dp_{4} \nonumber \\
= \frac{3}{2} -2p(1-p) -\frac{2}{3} p^{3} -2 (1-p+p^2) e^{-p}
+\frac{1}{2}(1-2p) e^{-2p} \nonumber \\ 
\eea


Consider Figure 3 next. It shows a singlet at site 2 followed by a doublet
at sites 4 and 5. An identical contribution will come from a situation
where the doublet occurs at sites 2 and 3, and the singlet at site 5 as
shown in Figure 4. We work out the contribution from Figure 3 explicitly,
and then multiply it by a factor 2 to take into account the contribution
from Figure 4. When site 2 flips up, site 3 will flip down if $h_{3} \le
h_{2}$, and if site 4 is up as well. Note that site 4 would have flipped
up earlier on ramp-II if $h_{4} \ge h_{5}$. If $h_{5} \ge h_{4}$, site 5
would have flipped up on ramp-II. In this case there is no possibility of
site 3 flipping down when site 2 flips up. The reason is as follows. The
presence of a doublet at sites 4 and 5 on plateau-I means that $h_{4} \le
h_{3}$ and $h_{5} \le h_{6}$. We must have $h_{3} \le h_{2}$ for the
creation of a new singlet (Rule-2). If $h_{4} \le h_{3}$, and $h_{3} \le
h_{2}$, then we have $h_{4} \le h_{2}$. Thus, when site 2 turns up on
ramp-III, site 4 will be down, and one of the two conditions for the
creation of a new singlet is not satisfied.

A similar reasoning as applied in the analysis of Figure 2 reveals that
site 3 must have flipped up after sites 1 and 6. The distribution of
fields at sites 2 and 5 in Figure 3 must be similar to the distribution of
fields at sites 2 and 4 in Figure 2. The distibution of fields at sites 3
and 4 must be uniform over the interval [$-\Delta$ to min($h_{1},h_{6}$)].
Thus the contributions of Figures 3 and 4 to the fraction of newly created
singlets when sites with $h_{i} \ge h$ have been relaxed on ramp-III is
given by,

\bea
P^{III}_{5} (p) = 2 \int_{0}^{p} \left[ 1-e^{-p_{2}} \right] dp_{2}
\int_{p_{2}}^{1}
dp_{3} \int_{p_{3}}^{1} dp_{4} \int_{p_{4}}^{1} \left[ 1 -e^{-p_{5}}
\right] dp_{5} \nonumber \\
=-\left(\frac{1}{3} + 3 e^{-1} \right)
+\left(\frac{1}{3} + 5 e^{-1} \right) p
-\left(\frac{1}{2} + 2 e^{-1} \right) p^{2}
+\frac{1}{3}\left( 1 + e^{-1} \right) p^{3} \nonumber \\ 
-\frac{1}{12} p^{4} 
+\left\{ \left(\frac{4}{3} + 3 e^{-1} \right)
-\left( 1 + 2 e^{-1} \right) p + e^{-1} p^{2}
-\frac{1}{3} p^{3} \right \} e^{-p} - e^{-2p} \nonumber \\
\eea


\section{Destruction of New Singlets}

The destruction of newly created singlets on ramp-III can be analysed in a
similar manner as their creation. For example, refer to Figure 2 again,
and recall that $h_{2} = \mbox{ min}(h_{2},h_{4})$, and $h_{3} \le h_{2}$.  
In this figure, the creation of new singlets when all sites with $h_{i}
\ge h$ have been relaxed is determined by $h_{2} \ge h$. The destruction
of new singlets when all sites with $h_{i} \ge h$ have been relaxed is
given by $h_{3} \ge h$. The result is an integral similar to the
expression for the creation of singlets. Only the limits of the
integration are altered. We obtain,

\bea
P^{III}_{6}(p) = 2 \int_{0}^{p} dp_{3} 
\int_{0}^{p_{3}} \left[ 1 -e^{-p_{2}} \right] dp_{2}
\int_{0}^{p_{3}} \left[ 1 -e^{-p_{4}} \right] dp_{4}
\nonumber \\
=\frac{1}{2} \left[ 1 -e^{-2p} \right] -2 p e^{-p} 
+ p (1-p) + \frac{1}{3} p^{3} \nonumber \\
\eea

Similarly, the contribution from Figures 3 and 4 is given by,

\bea
P^{III}_{7}(p) = 2 \int_{0}^{p} dp_{3} 
\int_{0}^{p_{3}}\left[ 1-e^{-p_{2}} \right] dp_{2}
\int_{p_{3}}^{1} dp_{4} 
\int_{p_{4}}^{1} \left[ 1 -e^{-p_{5}}
\right] dp_{5} \nonumber \\
= 2 e^{-1} - \left(1 + 4 e^{-1} \right) p
+\left(\frac{3}{2} + 3 e^{-1} \right) p^{2}
-\left( 1  + \frac{2}{3} e^{-1} \right) p^{3} +\frac{1}{4}p^{4}
\nonumber \\ 
- \left\{ \left( 1 + 2 e^{-1} \right)
- 2 \left( 1 +  e^{-1} \right) p + p^{2} \right \} e^{-p} + e^{-2p}
\nonumber \\ 
\eea


\section{Ramp-III}

Putting the various terms together, the probability that a randomly chosen
spin on the chain is up on ramp-III is given by

\bea
P_{\uparrow}^{III}(p) = P_{\uparrow}^{II}(1) + P_{1}^{III}(p) 
+ P_{2}^{III}(p) + P_{3}^{III}(p) \nonumber \\ 
- P_{4}^{III}(p) - P_{5}^{III}(p) + P_{6}^{III}(p) + P_{7}^{III}(p)
\eea

The magnetization on ramp-III is given by,
\be
m^{III}(h)=2 P_{\uparrow}^{III}(p) -1
\ee

After some simplification, we obtain

\bea
m^{III}(h) = - \left\{ \frac{13}{30} - \frac{53}{6} e^{-1} 
+ \frac{8}{3} e^{-2} \right\} 
+ \left\{ \frac{11}{6} - 18 e^{-1} + 8 e^{-2} \right\} p  \nonumber \\
-\left( 1 - 5 e^{-1} + 4 e^{-2} \right) p^{2}
+ \frac{1}{3} \left(1 + 2 e^{-1} +2 e^{-2} \right) p^{3}  \nonumber \\
+ \frac{1}{6} \left ( 1- 3 e^{-1} \right ) p^{4} 
+ \frac{1}{10} p^{5}  \nonumber \\
- \left\{ \left(\frac{8}{3} +10 e^{-1} \right)
-\left( 2 + 4 e^{-1} \right) p 
-\left( 4 - 2 e^{-1} \right) p^{2}
-\frac{2}{3} p^{3} \right \} e^{-p} \nonumber \\
+ \left(4+2 p)\right) e^{-2p} 
\eea

The above expression has been superposed on the simulation data in Figure
5. The agreement is excellent. The analytic result is indinguishable from
the simulation on the scale of Figure (5).

\section{Hysteresis Loop}

So far, we have analysed the magnetization $m(h_{a})$ in increasing
applied field. We have shown that the analytic result agrees with the
simulation rather well. The magnetization $m_R(h_a)$ in decreasing field
(return loop) is related to $m(h_a)$ by a symmetry of the model, i.e.
$m_R(h_a)=-m(-h_a)$. Thus we have implicitly determined the return loop as
well. The return loop has been shown in Figure (5) by a broken line. The
hysteresis in the anti-ferromagnetic RFIM is rather small, and the two
halves of the hysteresis loop lie very close to each other. In order to
show them more clearly, we have plotted in Figure (6) the separation
between the two halves of the hysteresis loop versus the applied field. To
be precise, we have plotted $\left[ m(h_a) -\overline{m} \right]$ and
$\left[ m_R(h_a) -\overline{m} \right]$ vs $h_{a}$, where $\overline{m}
=\frac{1}{2} \left[ m_R(h_a) + m(h_a) \right]$.  

As we may expect, the agreement between the theoretical expression and the
simulation is excellent on the scale of Figure 6 as well. In Figure 6 the
simulation data was obtained from a system of $10^{3}$ spins, and averaged
over $10^{3}$ independent realizations of the random field distribution.
The set of applied fields where the spins flip on each half of the
hysteresis loop is of course different for each realization of the random
field distribution. Therefore, the average over different realizations
requires a judgement on how to group the data. We divided the entire range
of the applied field from $[-2|J| -\Delta$ to $2|J| +\Delta]$ into $3
\times 10^{3}$ sections (bins)  of equal width. The data in each bin was
averaged separately. The simulation data shown in Figure 6 is a much
sparser set of data (in order not to crowd the figure). We have shown the
simulation data at intervals of $\delta h_{a}=.1$, and the theoretical
expression at intervals of $\delta h_{a}=.01$ (joined by a continuous line
on the lower half, and a broken line on the upper half of the hysteresis
loop).

\section{Discussion}

We have considered the zero-temperature dynamics of a one-dimensional
anti-ferromagnetic random field Ising model, and obtained an analytic
solution of the model if the following conditions apply:  
\begin{enumerate}

\item All spins are down initially, and the applied field is swept from
$h_{a}=-\infty$ to $h_{a}=+\infty$ infinitely slowly. The solution is also
applicable by symmetry to the case when all spins are up in the initial
state, and $h_{a}$ is decreased from $h_{a}=+\infty$ to $h_{a}=-\infty$.

\item The random field has a uniform bounded distribution in the interval
[$-\Delta$ to + $\Delta$]. We considered the case $\Delta \le |J|$. The
simplifying feature of this case is that the increasing applied field
exhausts all strings of down spins of length three or more (ramp-I) before
working on strings of down spins of length two (ramp-II). Similarly,
strings of down spins of length one (singlets) are turned up (ramp-III)  
only after the doublets are finished.

\end{enumerate}

The second condition mentioned above has been adopted essentially for
simplicity. It serves to illustrate the method of solution with a minimum
of algebraic detail. We do not see a conceptual difficulty in applying the
same method in the case $\Delta \ge |J|$, or when the random field has an
unbounded continuous distribution but we do not go into these details
here.

The restriction to an initial state where spins are either all down or all
up appears to be necessary so far. A similar difficulty is encountered in
the ferromagnetic random field Ising model~\cite{sanjib}. The relaxation
dynamics of the ferromagnetic model is qualitatively different from that
of the anti-ferromagnetic model. The relaxation process in the
ferromagnetic model is abelian, while in the anti-ferromagnetic model it
is not. We have made some progress in the methods of analytic solutions in
both cases, but a qualitatively new idea appears to be needed in extending
these methods to an arbitrary initial state.

One of us (PS) thanks D Dhar for several useful comments during the course
of this work.



\Large{Figure Captions}

\large{

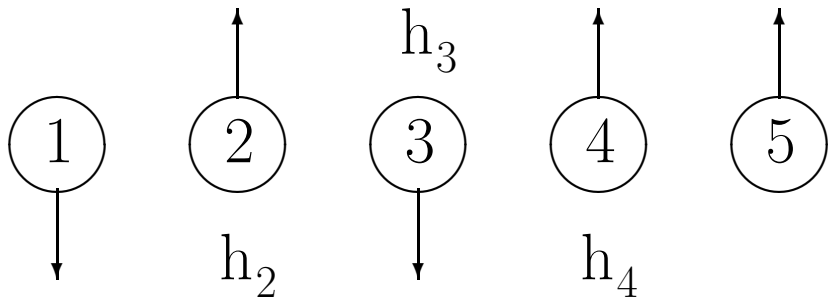
Figure 1: \\ A singlet (site 3) with one next nearest neighbor down (site
1), and one next nearest neighbor up (site 5). When the singlet turns up
at an applied field $h_{a}$, the spin at site 2 stays up if $\Delta \le
|J|$, but the spin at site 4 flips down if $h_{4} \le h_{3}$.

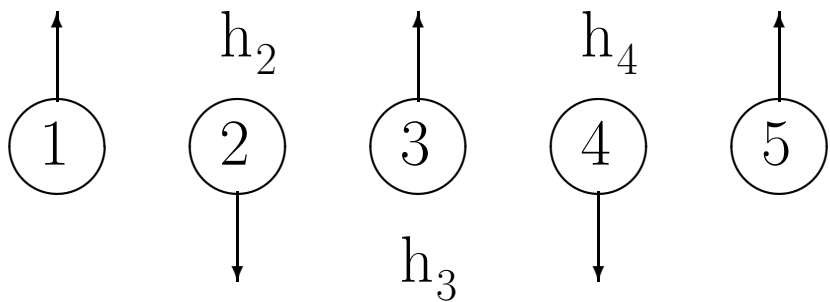
Figure 2: \\ Two adjacent singlets on Plateau-I: If $h_{2} =
\mbox{min}(h_{2},h_{4})$, and $h_{3} \le h_{2}$, then the spin at site 3
will flip down when the spin at site 2 flips up on ramp-III. This process
creates a new singlet on ramp-III.

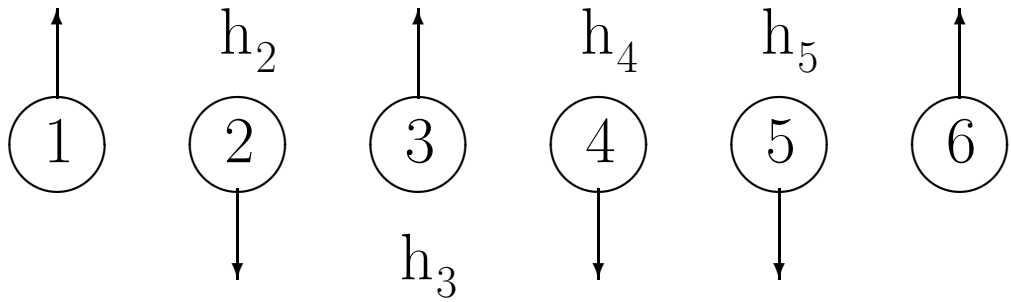
Figure 3: \\ A singlet followed by a doublet on Plateau-I: If $h_{4} \ge
h_{5}$, and $h_{3} \le h_{2}$, then a new singlet will be created at site
3 when the spin at site 2 turns up on ramp-III.

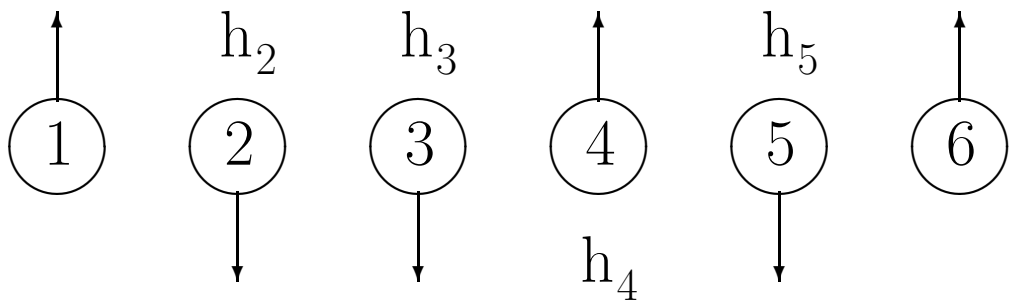
Figure 4: \\ A doublet followed by a singlet on Plateau-I: If $h_{3} \ge
h_{2}$, and $h_{4} \le h_{5}$, then a new singlet will be created at site
4 when the spin at site 5 turns up on ramp-III.

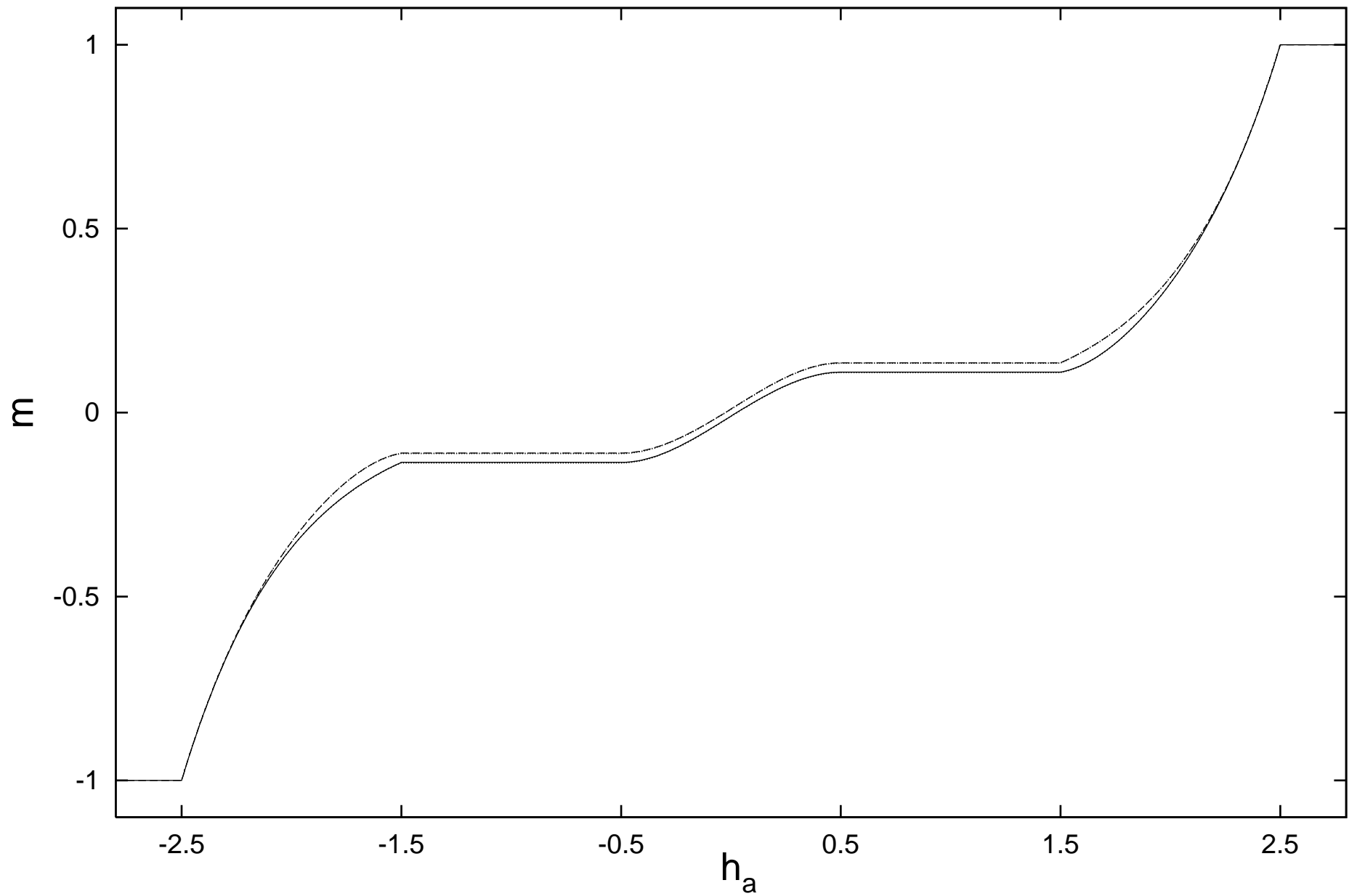
Figure 5: \\ Magnetization per spin $m$ in an applied field $h_{a}$ for an
anti-ferromagnetic RFIM ($J=-1$) for a rectangular distribution of the
random field of width $2\Delta=1$. The analytic result in increasing
applied field is shown by the solid line, and in decreasing field by a
broken line. Numerical result from a simulation of $10^{3}$ spins averaged
over $3 \times 10^{3}$ different realizations of the random field
distribution has been superposed on the theoretical result by dots. The
numerical result is indistinguishable from the theoretical result on the
scale of this figure.

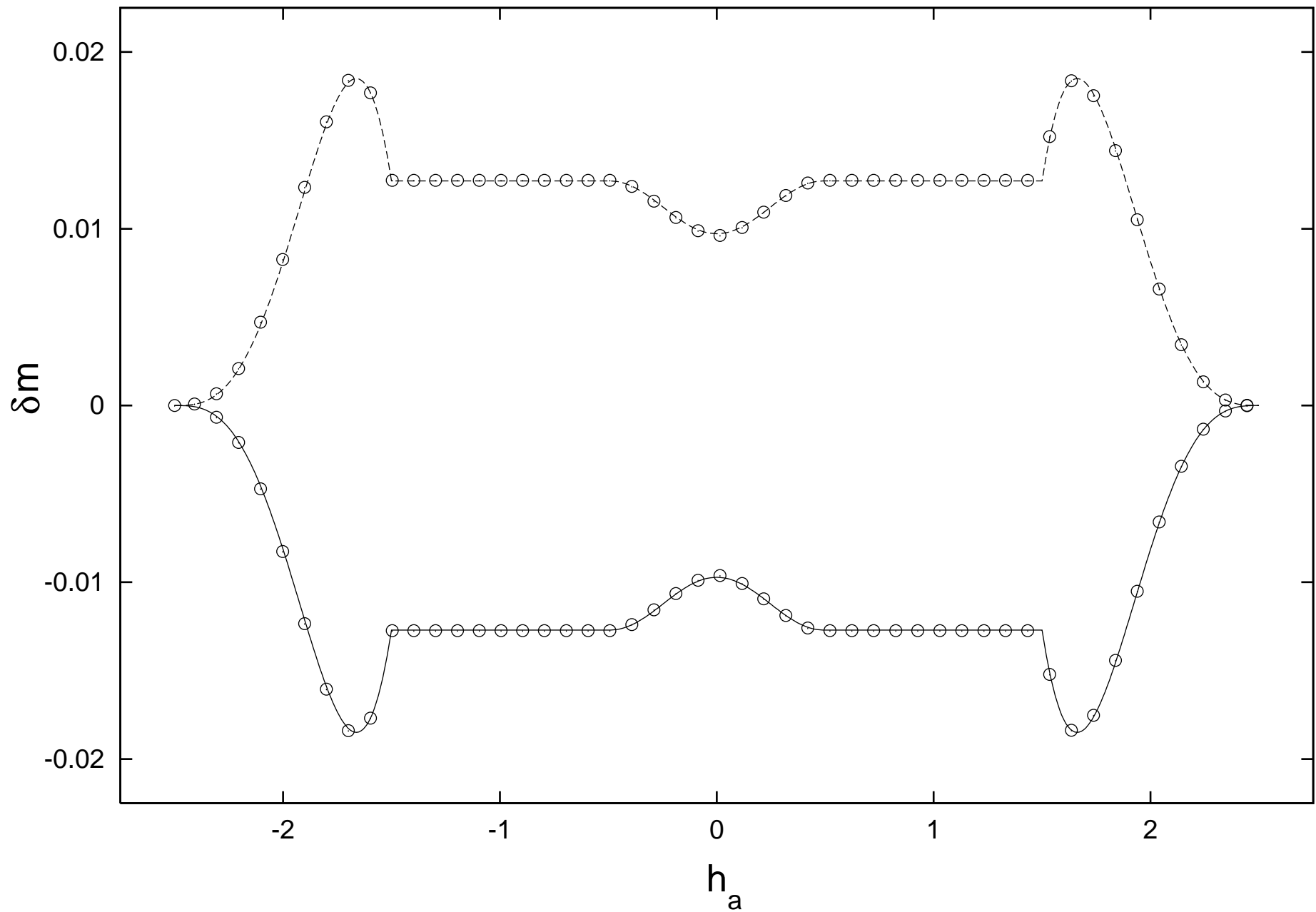
Figure 6: \\ Separation between the two halves of the hysteresis loop in
Figure 5 has been magnified by plotting it relative to the average value
of the two halves. The solid line shows $\delta m=\left[ m -\overline{m}
\right]$, and the broken line $\delta m = \left[ m_R -\overline{m}
\right]$ vs the applied field $h_a$; $\overline{m}=\frac{1}{2} \left[ m_R
+ m \right]$; m and $m_{R}$ are the magnetizations at applied field
$h_{a}$ in increasing and decreasing field respectively.

}


\newpage
\thispagestyle{empty}
\setlength{\unitlength}{1.2cm}
\begin{picture}(0.,0.)
\thicklines
\put(0,-2){\circle{1}}
\put(-.15,-2.2){\huge{1}} \put(0,-2.5){\vector(0,-1){1}}
\put(2,-2){\circle{1}}
\put(1.85,-2.2){\huge{2}} \put(2,-1.5){\vector(0,1){1}}
\put(4,-2){\circle{1}}
\put(3.85,-2.2){\huge{3}} \put(4,-2.5){\vector(0,-1){1}}
\put(6,-2){\circle{1}}
\put(5.85,-2.2){\huge{4}} \put(6,-1.5){\vector(0,1){1}}
\put(8,-2){\circle{1}}
\put(7.85,-2.2){\huge{5}} \put(8,-1.5){\vector(0,1){1}}

\put(1.8,-3.5){\huge{h}}
\put(2.2,-3.7){\bf{\Large{2}}}
\put(3.8,-1){\huge{h}}
\put(4.2,-1.2){\bf{\Large{3}}}
\put(5.8,-3.5){\huge{h}}
\put(6.2,-3.7){\bf{\Large{4}}}

\put(2.8,-6) {\bf{\huge{Figure 1}}}
\end{picture}

\setlength{\unitlength}{1.2cm}
\begin{picture}(-2,-8)
\thicklines
\put(0,-12){\circle{1}}
\put(2,-12){\circle{1}}
\put(4,-12){\circle{1}}
\put(6,-12){\circle{1}}
\put(8,-12){\circle{1}}
\put(0,-11.5){\vector(0,1){1}}
\put(2,-12.5){\vector(0,-1){1}}
\put(4,-11.5){\vector(0,1){1}}
\put(6,-12.5){\vector(0,-1){1}}
\put(8,-11.5){\vector(0,1){1}}
\put(-.2,-12.2){\huge{1}}
\put(1.8,-12.2){\huge{2}}
\put(3.8,-12.2){\huge{3}}
\put(5.8,-12.2){\huge{4}}
\put(7.8,-12.2){\huge{5}}
\put(1.8,-11){\huge{h}}
\put(2.2,-11.2){\bf{\Large{2}}}
\put(3.8,-13.5){\huge{h}}
\put(4.2,-13.7){\bf{\Large{3}}}
\put(5.8,-11){\huge{h}}
\put(6.2,-11.2){\bf{\Large{4}}}
\put(2.8,-16) {\bf{\huge{Figure 2}}}
\end{picture}

\newpage
\thispagestyle{empty}
\setlength{\unitlength}{1.2cm}
\begin{picture}(0.,0.)
\thicklines
\put(0,-2){\circle{1}}
\put(-.15,-2.2){\huge{1}} \put(0,-1.5){\vector(0,1){1}}
\put(2,-2){\circle{1}}
\put(1.85,-2.2){\huge{2}} \put(2,-2.5){\vector(0,-1){1}}
\put(4,-2){\circle{1}}
\put(3.85,-2.2){\huge{3}} \put(4,-1.5){\vector(0,1){1}}
\put(6,-2){\circle{1}}
\put(5.85,-2.2){\huge{4}} \put(6,-2.5){\vector(0,-1){1}}
\put(8,-2){\circle{1}}
\put(7.85,-2.2){\huge{5}} \put(8,-2.5){\vector(0,-1){1}}
\put(10,-2){\circle{1}}
\put(9.85,-2.2){\huge{6}} \put(10,-1.5){\vector(0,1){1}}

\put(1.8,-1){\huge{h}}
\put(2.2,-1.2){\bf{\Large{2}}}
\put(3.8,-3.5){\huge{h}}
\put(4.2,-3.7){\bf{\Large{3}}}
\put(5.8,-1){\huge{h}}
\put(6.2,-1.2){\bf{\Large{4}}}
\put(7.8,-1){\huge{h}}
\put(8.2,-1.2){\bf{\Large{5}}}

\put(2.8,-6) {\bf{\huge{Figure 3}}}
\end{picture}

\setlength{\unitlength}{1.2cm}
\begin{picture}(0.,0.)
\thicklines
\put(0,-12){\circle{1}}
\put(-.15,-12.2){\huge{1}} \put(0,-11.5){\vector(0,1){1}}
\put(2,-12){\circle{1}}
\put(1.85,-12.2){\huge{2}} \put(2,-12.5){\vector(0,-1){1}}
\put(4,-12){\circle{1}}
\put(3.85,-12.2){\huge{3}} \put(4,-12.5){\vector(0,-1){1}}
\put(6,-12){\circle{1}}
\put(5.85,-12.2){\huge{4}} \put(6,-11.5){\vector(0,1){1}}
\put(8,-12){\circle{1}}
\put(7.85,-12.2){\huge{5}} \put(8,-12.5){\vector(0,-1){1}}
\put(10,-12){\circle{1}}
\put(9.85,-12.2){\huge{6}} \put(10,-11.5){\vector(0,1){1}}

\put(1.8,-11){\huge{h}}
\put(2.2,-11.2){\bf{\Large{2}}}
\put(3.8,-11){\huge{h}}
\put(4.2,-11.2){\bf{\Large{3}}}
\put(5.8,-13.5){\huge{h}}
\put(6.2,-13.7){\bf{\Large{4}}}
\put(7.8,-11){\huge{h}}
\put(8.2,-11.2){\bf{\Large{5}}}

\put(2.8,-16) {\bf{\huge{Figure 4}}}
\end{picture}


\begin{thebibliography}{99}
\bibitem{ns} Although the new singlets make the analysis of ramp-III
somewhat tedious, but they save our model from an unphysical feature. If
we neglect the new singlets, the two halves of the hysteresis loop cross
each other twice so as to make a series of three subloops. In this case,
one could engineer a violation of the second law of thermodynamics by
running an engine over two of the subloops. However, the new singlets on
ramp-III push down the lower half of the hysteresis loop and prevent the
upper half from crossing it.
\end{thebibliography}
\end{document}